\documentclass[journal]{IEEEtran}
\usepackage[left=0.75in, right=0.75in, top=1in, bottom=0.9in]{geometry}
\usepackage{mathrsfs,amsthm}
\usepackage{cite}
\makeatletter
\def\ps@headings{%
\def\@oddhead{\mbox{}\scriptsize\rightmark \hfil \thepage}%
\def\@evenhead{\scriptsize\thepage \hfil \leftmark\mbox{}}%
\def\@oddfoot{}%
\def\@evenfoot{}}
\makeatother 
\usepackage{amsmath,amssymb,stfloats,subfigure,multicol}%amsmath
\usepackage{float}
\usepackage{graphicx}
\usepackage{epsfig,epsf,psfrag,amssymb,amsfonts,latexsym,graphicx,mathrsfs,subfigure}
\usepackage{bbm}
\usepackage{dsfont}
\usepackage{chngpage}
\usepackage[dvips]{color}
\usepackage{textcomp}
\usepackage{hyperref}
\usepackage{url}
\usepackage[hyphenbreaks]{breakurl}
\usepackage[normalem]{ulem}
\usepackage{algpseudocode}
\usepackage{caption}
\newsavebox{\ieeealgbox}

\usepackage{array}
\newtheorem{theorem}{Theorem}

\newtheorem{lemma}{Lemma}

\newtheorem{assumption}{Assumption}

 \def\old#1{}    % Please don't remove this... This command includes the text to be deleted.

\def\nn{\nonumber}
\def\beq{\begin{equation}}
\def\eeq{\end{equation}}
\def\bea{\begin{eqnarray}}
\def\eea{\end{eqnarray}}
\def\ba{\begin{array}}
\def\ea{\end{array}}

\def\bitem{\begin{itemize}}
\def\eitem{\end{itemize}}
\def\ben{\begin{enumerate}}
\def\een{\end{enumerate}}

\def\ie{{\it i.e.,\ \/}}

\definecolor{bgrd}{rgb}{1,1,1}
\definecolor{gray}{rgb}{0.5,0.5,0.5}
\definecolor{dkr}{rgb}{0.7,0.1,0.2}
\definecolor{dkb}{rgb}{0.1,0.1,0.8}

\makeatletter
\newdimen{\captionwidth}
\long\def\@makecaption#1#2{%
\captionwidth .9\hsize% use current value of \hsize
\vskip 10pt%
\setbox\@tempboxa\hbox{#1: #2}%
  \ifdim \wd\@tempboxa >\captionwidth%
    \setbox\@tempboxa\hbox{#1:\hspace*{.5em}}%
    \hfil\parbox{\captionwidth}{\raggedright\hangindent \wd\@tempboxa%
    \hangafter=1\unhbox\@tempboxa#2}\hfill%
  \else\centerline{\box\@tempboxa}%
  \fi
}
\makeatother
\def\scalefig#1{\epsfxsize #1\textwidth}

\def\edoc{\end{document}}

\def\mubf{\hbox{\boldmath$\mu$\unboldmath}}

\def\Thetabf{\hbox{$\bf \Theta$}}

\def\thetabf{{\mbox{\boldmath$\theta$\unboldmath}}}
\def\dbf{{\pmb d}}

\def\gbf{{\pmb g}}

\def\qbf{{\pmb q}}

\def\Ebf{{\pmb E}}

\def\Sbf{{\pmb S}}

\def\Ec{{\cal E}}

\begin{document}

\title{Convexifying Market Clearing of SoC-Dependent Bids from Merchant Storage Participants}
\author{Cong Chen,~\IEEEmembership{Student Member,~IEEE}
and~Lang~Tong,~\IEEEmembership{Fellow,~IEEE}% <-this % stops a space
% \thanks{\scriptsize Part of the work was presented at }
\thanks{\scriptsize
Cong Chen (\url{@cornell.edu}) and Lang Tong (\url{lt35@cornell.edu}) are with the Cornell University, Ithaca, NY 14853, USA. This work is supported in part by the National Science Foundation under Award 2218110 and 1932501, and in part by Power Systems and Engineering Research Center (PSERC) Research Project M-46.}}
\maketitle
%The work of Lang Tong and Cong Chen was supported in part by the National Science Foundation under Award under Grants 1809830 and 1932501, and in part by Power Systems and Engineering Research Center (PSERC) Research Project M-39

\begin{abstract}
State-of-charge (SoC) dependent bidding allows merchant storage participants to incorporate SoC-dependent operation and opportunity costs in a bid-based market clearing process. However,  such a bid results in a non-convex cost function in the multi-interval  economic  dispatch and market clearing, limiting its implementation in practice. We show that a simple restriction on the bidding format removes the non-convexity, making the multi-interval dispatch of SoC-dependent bids a standard convex piece-wise linear program. %\textcolor{red}{what's standard linear program? piece-wise linear is standard linear?}
\end{abstract}

\begin{IEEEkeywords}
Multi-interval economic dispatch, SoC dependent bid, convexification.
\end{IEEEkeywords}

\section{Introduction} \label{sec:intro}
Recent proposals \cite{CAISO_SOCdependent:22} have allowed merchant storage participants in the wholesale electricity market to submit state-of-charge (SoC) dependent offers and bids to capture more accurately the operation and opportunity costs of the energy storage  \cite{WangChen21hydroSoc, Ecker14JPSdegradaES, ZhengXu22impact}. With such bids, an economic dispatch program tends to schedule the battery SoC within a range favorable to the battery's health and the storage's ability to capture future opportunities under uncertainty.

However, a multi-interval economic dispatch with SoC-dependent bids involves integer variables \cite{ZhengXu22energy}, making the market clearing process computationally expensive for practical implementations. The nonconvexity of SoC-dependent bids also brings pricing challenges and the need for out-of-the-market uplift payments.

In this paper, we propose a simple restriction to the SoC-dependent bidding, referred to as the equal decremental-cost ratio (EDCR) condition, that transforms the nonconvex economic dispatch optimization into a convex piece-wise linear program compatible with the standard market clearing process. A procedure to produce bids satisfying the EDCR condition from the true bid-in cost functions is also proposed.

\section{SoC-dependent bid and dispatch models} \label{sec:model}
\subsection{Storage and SoC-dependent cost models}

We assume the standard imperfect storage model.  In the scheduling interval $t$, let  $e_t$ be the storage SoC, $g_t^{\mbox{\tiny C}}$ the charging power, and $g_t^{\mbox{\tiny D}}$ the discharging power, respectively. The storage SoC evolves according to 
\beq
\begin{array}{lcl}\label{eq:SoCevolve}
e_{t+1}=e_{t} +g^{\mbox{\tiny C}}_{t}\eta^{\mbox{\tiny C}}-g^{\mbox{\tiny D}}_{t}/\eta^{\mbox{\tiny D}},~~ g^{\mbox{\tiny C}}_{t}g^{\mbox{\tiny D}}_{t}=0,
\end{array}
\eeq
where $\eta^{\mbox{\tiny C}},\eta^{\mbox{\tiny D}} \in (0,1]$ are charging/discharging efficiencies.
\vspace{-0.4cm}

\begin{figure}[h]
\center
\begin{psfrags}
\scalefig{0.25}\epsfbox{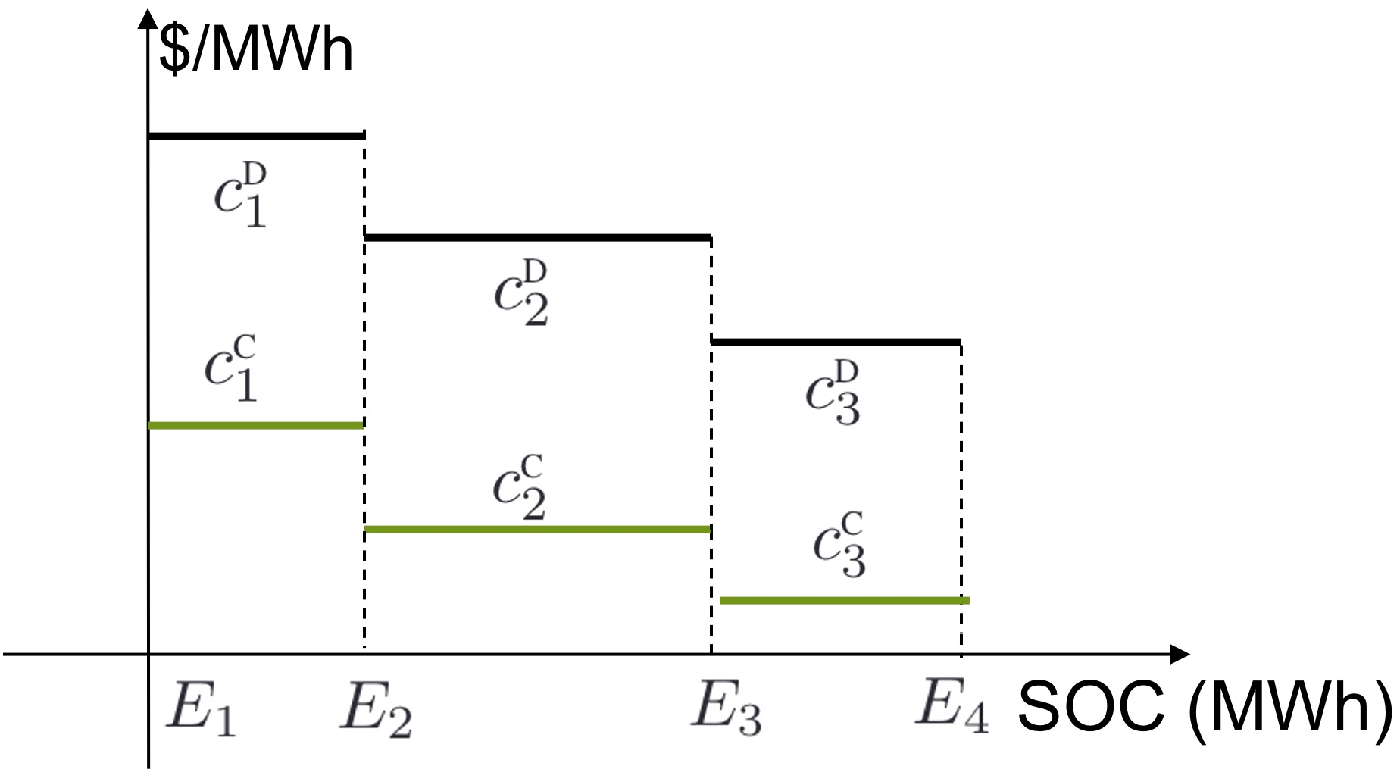}\scalefig{0.25}\epsfbox{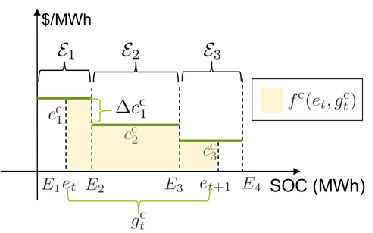}
\end{psfrags}
\vspace{-2em}\caption{\scriptsize  Left: The SoC-dependent bid and offer format when $K=3$. Right: Cost of charging the storage by $g^{\mbox{\tiny C}}_{t} $ from $e_t$ to $e_{t+1}$. }
\label{fig:SOC_D_Bid}
\end{figure}
\vspace{-0.3cm}

A standard piecewise-linear SoC-dependent bid model \cite{CAISO_SOCdependent:22} is illustrated in Fig.~\ref{fig:SOC_D_Bid} (left). Without loss of generality, we partition the SoC axis into $K$ consecutive segments, within each segment $\Ec_k:=[E_k,E_{k+1}]$, a pair of bid-in {\em marginal cost/benefit} parameters  $(c_k^{\mbox{\tiny C}},c_k^{\mbox{\tiny D}})$ is defined. The marginal discharging (bid-in) costs (to the grid) $b^{\mbox{\tiny D}}(e_t; \mathbf{c}^{\mbox{\tiny D}},\mathbf{E})$ and  marginal charging (bid-in)  benefits (from the grid) $b^{\mbox{\tiny C}}(e_t; \mathbf{c}^{\mbox{\tiny C}},\mathbf{E})$ are functions of battery SoC $e_t$. In particular, using the indicator function\footnote{$\mathbbm{1}_{\{s \in \Ec_i\}}$ equals to 1 when $s \in \Ec_i$.}  $\mathbbm{1}$,  
\vspace{-0.2cm}
\beq\label{eq:SoCBid}
\begin{array}{r}
%b^{\mbox{\tiny C}}(e_t; \mathbf{c}^{\mbox{\tiny C}},\mathbf{E})=\sum_{k=1}^Kc^{\mbox{\tiny C}}_k\mathbf{1}_{\{e_t\in \Ec_k\}},~b^{\mbox{\tiny D}}(e_t; \mathbf{c}^{\mbox{\tiny D}},\mathbf{E})=\sum_{k=1}^Kc^{\mbox{\tiny D}}_k\mathbf{1}_{\{e_t\in \Ec_k\}}
\left\{\begin{array}{ll}
b^{\mbox{\tiny C}}(e_t; \mathbf{c}^{\mbox{\tiny C}},\mathbf{E}):=\sum_{k=1}^Kc^{\mbox{\tiny C}}_k\mathbbm{1}_{\{e_t\in \Ec_k\}}\\
b^{\mbox{\tiny D}}(e_t; \mathbf{c}^{\mbox{\tiny D}},\mathbf{E}):=\sum_{k=1}^Kc^{\mbox{\tiny D}}_k\mathbbm{1}_{\{e_t\in \Ec_k\}}
\end{array}\right.
\end{array}
\eeq\vspace{-0.1cm}
with $\mathbf{c}^{\mbox{\tiny C}}:=(c_k^{\mbox{\tiny C}})$, $\mathbf{c}^{\mbox{\tiny D}}:=(c_k^{\mbox{\tiny D}})$ and $\mathbf{E}:= (E_k)$ as parameters.

For the longevity of the battery and the ability to capture profit opportunities, it is more costly to discharge when the SoC is low, and the benefit of charging is small when the SoC is high. Therefore, typical  bid-in discharge costs $(c^{\mbox{\tiny D}}_{k})$ and charging benefits $(c^{\mbox{\tiny C}}_{k})$ are monotonically decreasing. Furthermore,  the storage participant is willing to discharge only if the selling price is higher than the buying price. Hence, the storage participant's willingness to sell by discharge (adjusted to the discharging efficiency) must be higher than its willingnesss to purchase (adjusted to the charging efficiency),  \ie $c^{\mbox{\tiny D}}_{\mbox {\tiny K}}\eta^{\mbox{\tiny D}} > c^{\mbox{\tiny C}}_{1}/\eta^{\mbox{\tiny C}}$.  Together,  SoC-dependent bids and offers satisfy the following.

%Besides, from the perspective of a storage participant, the charging/discharging costs are discounted by corresponding efficiency parameters, and the compensation from discharging to the grid should be greater than the marginal benefit of charging, \ie $c^{\mbox{\tiny D}}_{k}\eta^{\mbox{\tiny D}} > c^{\mbox{\tiny C}}_{k}/\eta^{\mbox{\tiny C}}, \forall k$ at all SoC levels. Therefore, It is reasonable to assume that the {\em SoC-dependent bid parameters}  satisfy the following assumption.
\begin{assumption}[] \label{assume:single} The SoC-dependent cost/benefit parameters $\{(c_k^{\mbox{\tiny C}},c_k^{\mbox{\tiny D}}), \eta^{\mbox{\tiny C}}, \eta^{\mbox{\tiny D}}\}$ satisfy the following monotonicity conditions $\forall k=1,\cdots, K-1$:
\[
\left\{\begin{array}{l}
c_k^{\mbox{\tiny C}} \ge c_{k+1}^{\mbox{\tiny C}}\\
c_k^{\mbox{\tiny D}} \ge c^{\mbox{\tiny D}}_{k+1}\\
\end{array}\right.~~{\rm and}~~ c^{\mbox{\tiny C}}_{1}/\eta^{\mbox{\tiny C}} < c^{\mbox{\tiny D}}_{\mbox {\tiny K}}\eta^{\mbox{\tiny D}}.\]
%\underset{k}{\rm max}{~c^{\mbox{\tiny C}}_{k}/\eta^{\mbox{\tiny C}}} < \underset{k}{\rm min}{~c^{\mbox{\tiny D}}_{k}\eta^{\mbox{\tiny D}}}.\]
\end{assumption}

\subsection{Cost function of SoC-dependent bids}

SoC-dependent bids and offers induce SoC-dependent scheduling costs involving the ({\it ex ante}) SoC $e_t$ in scheduling stage $t$ before the dispatch and the ({\it ex post}) SoC $e_{t+1}$ after the dispatch that may be in a different SoC partitioned segment.  Specifically, the stage cost $f(g^{\mbox{\tiny C}}_t, g^{\mbox{\tiny D}}_t, e_t)$ in interval $t$ is given by\footnote{For simplicity, indexes and ramping costs for storage are ignored here.} 
\beq\label{eq:StageCost}
f(g^{\mbox{\tiny C}}_t, g^{\mbox{\tiny D}}_t, e_t):= f^{\mbox{\tiny D}}(e_t,g_t^{\mbox{\tiny D}})-f^{\mbox{\tiny C}}(e_t,g_t^{\mbox{\tiny C}}),
\eeq
where $f^{\mbox{\tiny D}}$ is the discharging cost, and $f^{\mbox{\tiny C}}$ is the charging benefit. In particular, for every $e_t \in \Ec_m$ and $e_{t+1} \in \Ec_{n}$,  %$f^{\mbox{\tiny C}}(\cdot)$ and $f^{\mbox{\tiny D}}(\cdot)$ are respectively defined as follows: 

%\beq\label{eq:SoCBid}
%\begin{array}{c}
\[f^{\mbox{\tiny C}}(e_t, g_t^{\mbox{\tiny C}}) := \mathbbm{1}_{\{n\geq m\}}g^{\mbox{\tiny C}}_tc^{\mbox{\tiny C}}_{n}+\mathbbm{1}_{\{n > m\}}\sum_{k=m}^{n-1}\frac{\Delta c^{\mbox{\tiny C}}_{k}}{\eta^{\mbox{\tiny C}}}(E_{k+1}-e_t),\]
%\[f^{\mbox{\tiny C}}(e_t, g_t^{\mbox{\tiny C}}) = 
%\left\{\begin{array}{ll}
% g^{\mbox{\tiny C}}_tc^{\mbox{\tiny C}}_{n}   & m=n\\
%  g^{\mbox{\tiny C}}_tc^{\mbox{\tiny C}}_{n} +\sum_{k=m}^{n-1}\frac{\Delta c^{\mbox{\tiny C}}_{k}}{\eta^{\mbox{\tiny C}}}(E_{k+1}-e_t) & n>m\\% 0 & \mbox{otherwise}
% \end{array}\right.\]
\[f^{\mbox{\tiny D}}(e_t, g_t^{\mbox{\tiny D}}): =\mathbbm{1}_{\{n\le m\}}g^{\mbox{\tiny D}}_tc^{\mbox{\tiny D}}_{n}+\mathbbm{1}_{\{n < m\}}\sum_{k=n+1}^{m}\eta^{\mbox{\tiny D}}\Delta c^{\mbox{\tiny D}}_{k-1}(E_{k}-e_t)\]
%\end{array}
%\eeq
%\[f^{\mbox{\tiny D}}(e_t, g_t^{\mbox{\tiny D}}) = 
%\left\{\begin{array}{ll}
% g^{\mbox{\tiny D}}_tc^{\mbox{\tiny D}}_{n}   &  m=n\\
% g^{\mbox{\tiny D}}_tc^{\mbox{\tiny D}}_{n} +\sum_{k=n+1}^{m}\eta^{\mbox{\tiny D}}\Delta c^{\mbox{\tiny D}}_{k}(E_{k}-e_t) & n<m\\
%0 & \mbox{otherwise}
%\end{array}\right.
%\]
with $\Delta c^{\mbox{\tiny C}}_{k}:=c^{\mbox{\tiny C}}_{k}-c^{\mbox{\tiny C}}_{k+1}$ and $\Delta c^{\mbox{\tiny D}}_{k}:=c^{\mbox{\tiny D}}_{k}-c^{\mbox{\tiny D}}_{k+1}$. Fig.~\ref{fig:SOC_D_Bid} (right)  illustrates $f^{\mbox{\tiny C}}(e_t, g_t^{\mbox{\tiny C}})$ in an example with $K=3, m=1$, and  $n=3$.  Note that the stage cost  $f(g^{\mbox{\tiny C}}_t, g^{\mbox{\tiny D}}_t, e_t)$ is nonconvex, although it is convex if given $e_t$.  

 \subsection{The multi-interval economic dispatch} 
We consider a multi-interval dispatch model involving $T$ intervals and $M$ buses.  In decision interval $t$, let  $g_{it}^{\mbox{\tiny C}}$ and $g_{it}^{\mbox{\tiny D}}$ be the charging and discharging decision variables, respectively, and let $e_{it}$ be the SoC of unit $i$. With the single stage cost in (\ref{eq:StageCost}), the $T$-interval  operation cost of storage $i$ is given by
  \beq\label{eq:EScostMulti}
\begin{array}{l}
F_i(\gbf^{\mbox{\tiny C}}_i,\gbf^{\mbox{\tiny D}}_i; s_i):=\sum_{t=1}^{T}f_i(g^{\mbox{\tiny C}}_{it}, g^{\mbox{\tiny D}}_{it}, e_{it}),
\end{array}
\eeq 
where $\gbf^{\mbox{\tiny C}}_i, \gbf^{\mbox{\tiny D}}_i \in \mathbb{R}^T$ denote the vector of charging and discharging power for storage $i$ over $T$-interval, respectively. 
 
For the interval $t$, let $d_{it}$ be the demand at bus $i$ and $\dbf[t]:=(d_{1t},\cdots, d_{Mt})$ the demand vector for all buses.  Let $\gbf^{\mbox{\tiny G}}[t]:=(g^{\mbox{\tiny G}}_{1t},\cdots, g^{\mbox{\tiny G}}_{Mt})$ be the vector of bus generations. Similarly defined are $\gbf^{\mbox{\tiny D}}[t]$ and $\gbf^{\mbox{\tiny C}}[t]$ as the vector of charging and discharging power of the battery storage, respectively. For simplicity, we establish the dispatch model with one generator and one storage at each bus, which is extendable to general cases. Given the convex generator cost $f^{\mbox{\tiny G}}_i(g^{\mbox{\tiny G}}_{it})$, the initial SoC $e_{i1}=s_i$, and the load forecast $(\dbf[t])$ over the  $T$-interval scheduling horizon, the economic dispatch minimizes the system operation costs is given by % In a rolling-window dispatch framework, only the first interval decision is realized.  
\beq \label{eq:NONCVX}
\begin{array}{lrl}
&\underset{\{(g_{it}^{\mbox{\tiny G}},g_{it}^{\mbox{\tiny C}}, g_{it}^{\mbox{\tiny D}}, e_{it})\}}{\rm minimize} & \sum_{i=1}^{M}(F_i(\gbf^{\mbox{\tiny C}}_i,\gbf^{\mbox{\tiny D}}_i; s_i)+\sum_{t=1}^{T}f^{\mbox{\tiny G}}_i(g^{\mbox{\tiny G}}_{it}))\\
%\Big(F(\gbf^c,\gbf^d,\sbf):=\sum_{t=1}^T\varphi(\gbf^{\mbox{\tiny G}}[t],\gbf^{\mbox{\tiny C}}[t], \gbf^{\mbox{\tiny D}}[t])\Big)\\
& \mbox{subject to}& \forall t\in \{1,...,T\}, \forall i\in \{1,...,M\}\\
&\pmb{\mubf}[t]: & \Sbf (\gbf^{\mbox{\tiny G}}[t]+\gbf^{\mbox{\tiny D}}[t]-{\gbf}^{\mbox{\tiny C}}[t]-\dbf[t]) \le \qbf\\
&\lambda_{t}:& {\bf 1}^\intercal (\gbf^{\mbox{\tiny G}}[t]+\gbf^{\mbox{\tiny D}}]t]-\gbf^{\mbox{\tiny C}}[t])={\bf 1}^\intercal \dbf[t]\\
% & & \hfill \mbox{for all $t\le t'\ < t+W.$}\\[0.2em]
% &&E_{s_{T+1}}\le e_{T+1}\le E_{s_{T+1}+1},\\
&\phi_{it}:& e_{it}+g^{\mbox{\tiny C}}_{it}\eta^{\mbox{\tiny C}}-g^{\mbox{\tiny D}}_{it}/\eta^{\mbox{\tiny D}}=e_{i(t+1)}\\
&(\underline{\rho}^{\mbox{\tiny G}}_{it},\bar{\rho}^{\mbox{\tiny G}}_{it}):& 0 \le g^{\mbox{\tiny G}}_{it}\le \bar{g}^{\mbox{\tiny G}}_i\\
&(\underline{\rho}^{\mbox{\tiny C}}_{it},\bar{\rho}^{\mbox{\tiny C}}_{it}):& 0 \le g^{\mbox{\tiny C}}_{it}\le \bar{g}^{\mbox{\tiny C}}_i\\
&(\underline{\rho}^{\mbox{\tiny D}}_{it},\bar{\rho}^{\mbox{\tiny D}}_{it}):& 0 \le g^{\mbox{\tiny D}}_{it}\le \bar{g}^{\mbox{\tiny D}}_i\\
&&\underline{E}_i \le e_{i(t+1)}\le \bar{E}_i\\
&&g^{\mbox{\tiny C}}_{it}g^{\mbox{\tiny D}}_{it}=0\\
&&e_{i1}=s_i,
\end{array}
\eeq
where the DC power flow model is considered with the shift-factor matrix  $\Sbf \in \mathbb{R}^{2B\times M}$ for a network with $B$ branches and the branch flow limit $ \qbf \in \mathbb{R}^{2B}$. The system operation constraints include power balance constraints, SoC state-transition constraints, charging/discharging capacity limits, and SoC limits. The bilinear constraint, $g^{\mbox{\tiny C}}_{it}g^{\mbox{\tiny D}}_{it}=0, \forall i, t$, prevents the simultaneous charging and discharging decisions. %With the optimal dual solutions, \ie $\lambda^*_t \in \mathbb{R}, \mubf^*[t] \in \mathbb{R}^{2B}$, the locational marginal price (LMP) in interval $t$ at bus $i$ is $\pi_{it}:=\lambda^*_t-\Sbf(:,i)^\intercal\mubf^*[t]$.

Note that (\ref{eq:NONCVX}) is nonconvex for two reasons. First, the objective function is nonconvex and subdifferentiable because the nonconvex multi-stage storage operation cost (\ref{eq:EScostMulti}), as is shown in Fig.~\ref{fig:SOCdependent} (top left). Second, the equality constraint banning  simultaneous charging/discharging decisions in  (\ref{eq:NONCVX}) is bilinear. In the following section, we remove these two forms of nonconvexities. %by imposing a condition on the bid-in cost function under the assumption of nonnegative LMPs. 

\section{Convexifying Market Clearing} \label{sec:Condi}
We now convexify the objective function and relax the bilinear equality constraints of the market clearing problem (\ref{eq:NONCVX}). Theorem~\ref{thm:Eq} below gives a condition on the bid-in cost parameters that convexify  the objection function\footnote{Storage index $i$ is omitted in Theorem~\ref{thm:Eq} and Sec.~\ref{sec:OptApprox} for simplicity.}.  

% From the above analysis, we know that Assumption~\ref{assume:single} and conditions in Lemma~\ref{lemma:bidSpread} are not enough to guarantee a convex dispatch problem (\ref{eq:NONCVX}) in multi-interval dispatch. Here we propose  a slight restriction of the biding format to remove non-convexity.
\begin{theorem}[] \label{thm:Eq}
If a storage participant's bid-in parameters satisfy the equal decremental-cost ratio (EDCR) condition,
\beq\label{eq:EDCR}
\frac{c^{\mbox{\tiny C}}_k-c^{\mbox{\tiny C}}_{k-1}}{c^{\mbox{\tiny D}}_k-c^{\mbox{\tiny D}}_{k-1}}=\eta^{\mbox{\tiny C}}\eta^{\mbox{\tiny D}}, \forall k, 
\eeq
under Assumption~\ref{assume:single}, the multi-interval storage operation cost in (\ref{eq:EScostMulti}) is piecewise linear convex given by
%the total storage operation cost over multi-interval, given initial SoC $e_1=s$, is a piecewise linear convex function of $\gbf^{\mbox{\tiny C}}$ and $\gbf^{\mbox{\tiny D}} $, given by % \textcolor{red}{$\alpha_0$ linear form, horizontal check? Lebesgue integral}%$F(g^{\mbox{\tiny C}}_t, g^{\mbox{\tiny D}}_t;s)=\sum_{t=1}^{T}f(g^{\mbox{\tiny C}}_{t}, g^{\mbox{\tiny D}}_{t};e_{t})$
\beq
\begin{array}{lrl}\label{eq:ES_cost}
F(\gbf^{\mbox{\tiny C}}, \gbf^{\mbox{\tiny D}};s)=\underset{j\in\{1,...,K\}}{\rm max}\{\alpha_j(s)-c^{\mbox{\tiny C}}_{j}\mathbf{1}^\intercal \mathbf{g}^{\mbox{\tiny C}}+c^{\mbox{\tiny D}}_{j}\mathbf{1}^\intercal \mathbf{g}^{\mbox{\tiny D}}\}
%F(\gbf^{\mbox{\tiny C}}, \gbf^{\mbox{\tiny D}};s)=-c^{\mbox{\tiny C}}_{n}\mathbf{1}^\intercal \mathbf{g}^{\mbox{\tiny C}}+c^{\mbox{\tiny D}}_{n}\mathbf{1}^\intercal \mathbf{g}^{\mbox{\tiny D}}\\
%~~~~~~~~+\begin{cases}
%\sum_{k=m}^{n-1}\frac{-\Delta c^{\mbox{\tiny C}}_{k}}{\eta^{\mbox{\tiny C}}}(E_{k+1}-s), & n>m \\
%0, & m=n  \\
%\sum_{k=n+1}^{m}\eta^{\mbox{\tiny D}}\Delta c^{\mbox{\tiny D}}_{k-1}(E_{k}-s), & n<m
%\end{cases},
\end{array}
\eeq
with $\alpha_j(s):=-\sum_{k=1}^{j-1}\frac{\Delta c^{\mbox{\tiny C}}_k(E_{k+1}-E_1)}{\eta^{\mbox{\tiny C}}}-\frac{c^{\mbox{\tiny C}}_{j}(s-E_1)}{\eta^{\mbox{\tiny C}}}+h(s)$ and $h(s):=\sum_{i=1}^K\mathbbm{1}_{\{s\in \Ec_i\}}(\frac{c^{\mbox{\tiny C}}_i(s-E_1)}{\eta^{\mbox{\tiny C}}}+\sum_{k=1}^{i-1}\frac{\Delta c^{\mbox{\tiny C}}_k (E_{k+1}-E_1)}{\eta^{\mbox{\tiny C}}})$. %$m, n$ are indexes for SoC-partitioned sets that have $e_1 \in \Ec_m, e_{\mbox{\tiny T+1}}\in \Ec_{n}$, and
\end{theorem}

The proof is given in the appendix. Note that if bid-in costs are derived from the value function of the stochastic storage optimization based on price forecasts as in  \cite{ZhengXu22impact, ZhengXu22energy}, the derived bids satisfy (\ref{eq:EDCR})\footnote{Adopting SoC-independent marginal discharge cost and efficiency parameters as used in \cite{ZhengXu22impact}, the SoC-dependent bid derived in equation (4) of \cite{ZhengXu22energy} satisfies the EDCR condition  (\ref{eq:EDCR}) in this paper.}. The following lemma supports the exact relaxation of $g^{\mbox{\tiny C}}_{it}g^{\mbox{\tiny D}}_{it}=0, \forall i,t$. %by showing the exact relaxation with non-negative LMPs \footnote{Negative LMP is considered in \cite{ChenBaldick21TPSstorageSCUCBinary} with a mixed integer storage model.}. 
%With the EDCR condition satisfied, the economic dispatch problem (\ref{eq:NONCVX}) has a convex and subdifferentiable objective function. 

\begin{lemma}[] \label{lemma:bidSpread}
Under Assumption~\ref{assume:single} and EDCR condition, if the locational marginal prices (LMPs) from the relaxed economic dispatch are non-negative, the relaxation of the bilinear constraints $g^{\mbox{\tiny C}}_{it}g^{\mbox{\tiny D}}_{it}=0, \forall i, t$ in  (\ref{eq:NONCVX}) is exact. 
\end{lemma}

The proof is given in the appendix. The computation of LMP (after relaxation of the bilinear constraint) is standard.  Specifically, the LMP for bus $i$ and interval $t$ is defined by $\pi_{it}:=\lambda^*_t-\Sbf(:,i)^\intercal\mubf^*[t]$ with the optimal dual solutions of (\ref{eq:NONCVX}) after relaxing the bilinear equality  constraints.%,  \ie $\lambda^*_t \in \mathbb{R}, \mubf^*[t] \in \mathbb{R}^{2B}$, after relaxing the bilinear constraints. Besides, the negative LMP is considered in \cite{ChenBaldick21TPSstorageSCUCBinary} with a mixed integer storage model. 

The non-negative assumption on LMP has been considered in \cite{Li18CSEEstorage, ChenBaldick21TPSstorageSCUCBinary} for the exact relaxation of bilinear constraint in (5) for differentiable objective functions.  Here we have a slight generalization for a convex piecewise linear objective function by deploying the subgradient measure  \cite[p. 281]{Rockafellar70convex}. See the proof in Appendix.  

%Such an exact relaxation is shown under the assumption of a convex and differentiable objective function in  \cite{Li18CSEEstorage, ChenBaldick21TPSstorageSCUCBinary}. By imposing the EDCR constraint, Theorem~\ref{thm:Eq} guarantees that the cost function of (\ref{eq:NONCVX}) is convex, but not necessarily differentiable.  Here we remove the assumption of differentiability and deploy the subgradient measure  \cite[p. 281]{Rockafellar70convex} to deal with discontinuities at the boundaries of SoC segments, when proving  Lemma~\ref{lemma:bidSpread} in the appendix .  

\section{Optimal EDCR approximatkon} \label{sec:OptApprox}
%\input OptEDCR_v1
%\subsection{Appriximate bidding curve directly}
In constructing the SoC-dependent storage bids and offers in (\ref{eq:SoCBid}), the true marginal costs (or true marginal cost $\tilde{b}^{\mbox{\tiny D}}(e_t)$ and marginal benefit  $\tilde{b}^{\mbox{\tiny C}}(e_t)$) may not satisfy the EDCR condition. The following optimization aims at finding the optimal approximation of $\tilde{b}^{\mbox{\tiny C}}(e_t)$ and $\tilde{b}^{\mbox{\tiny D}}(e_t)$ with the EDCR condition satisfied by parameters $\thetabf=\{\mathbf{c}^{\mbox{\tiny C}}, \mathbf{c}^{\mbox{\tiny D}}, \Ebf\}$,   % \in \mathbb{R}^{3K-1}
\beq \label{eq:OptEDCR_Bid}
\begin{array}{lrl}
&\underset{\thetabf \in \Thetabf}{\rm minimize} &||b^{\mbox{\tiny C}}(\cdot|\thetabf)-\tilde{b}^{\mbox{\tiny C}}(\cdot)||_2^2+||b^{\mbox{\tiny D}}(\cdot|\thetabf)-\tilde{b}^{\mbox{\tiny D}}(\cdot)||_2^2.\\
%& s.t.& \forall k\in \{1,...,K\}\\
%&&c_k^{\mbox{\tiny C}} \ge c_{k+1}^{\mbox{\tiny C}},\\
% &&E_{s_{T+1}}\le e_{T+1}\le E_{s_{T+1}+1},\\
%&& c_k^{\mbox{\tiny D}} \ge c_{k+1}^{\mbox{\tiny D}},\\
%&&c^{\mbox{\tiny C}}_{1}/\eta^{\mbox{\tiny C}} < c^{\mbox{\tiny D}}_{\mbox {\tiny K}}\eta^{\mbox{\tiny D}},\\
%&&(\tilde{c}^{\mbox{\tiny D}}_k-\tilde{c}^{\mbox{\tiny D}}_{k-1})\eta^{\mbox{\tiny C}}\eta^{\mbox{\tiny D}}=\tilde{c}^{\mbox{\tiny C}}_k-\tilde{c}^{\mbox{\tiny C}}_{k-1},
\end{array}
\eeq

The objective fuction measures the distance between the original {\em true marginal cost} and the approximation bids/offers, and $\thetabf$ is restricted in a set $\Thetabf$ satisfying Assumption~\ref{assume:single}  and the EDCR condition from Theorem~\ref{thm:Eq}. With $N$ data samples $(S_n, B^{\mbox{\tiny C}}_n, B^{\mbox{\tiny D}}_n)_{n=1}^N$ from the  {\em true marginal cost}, the objective is $\frac{1}{N}\sum_{n=1}^N((b^{\mbox{\tiny C}}(S_n|\thetabf)-B^{\mbox{\tiny C}}_n)^2+(b^{\mbox{\tiny D}}(S_n|\thetabf)-B^{\mbox{\tiny D}}_n)^2)$. 

Optimization (\ref{eq:OptEDCR_Bid}) for the optimal EDCR approximation is in general nonconvex. By fixing $\Ebf$ while solving for $(\mathbf{c}^{\mbox{\tiny C}}, \mathbf{c}^{\mbox{\tiny D}})$, or fixing $(\mathbf{c}^{\mbox{\tiny C}}, \mathbf{c}^{\mbox{\tiny D}})$ while solving for $\Ebf$, we can iteratively approach the (local) optimal solution by solving a convex problem in each iteration.

\section{Example} \label{sec:Ex}

\begin{figure}[h]
\center
\begin{psfrags}
\scalefig{0.24}\epsfbox{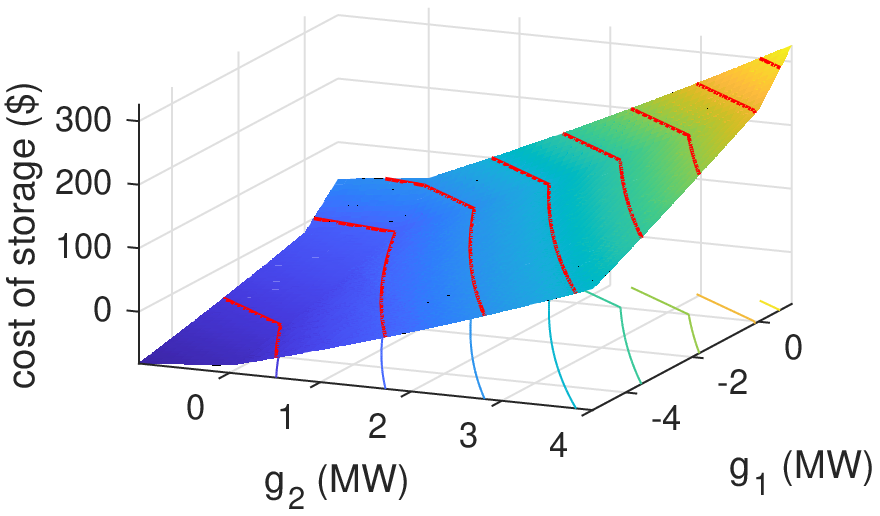}\scalefig{0.24}\epsfbox{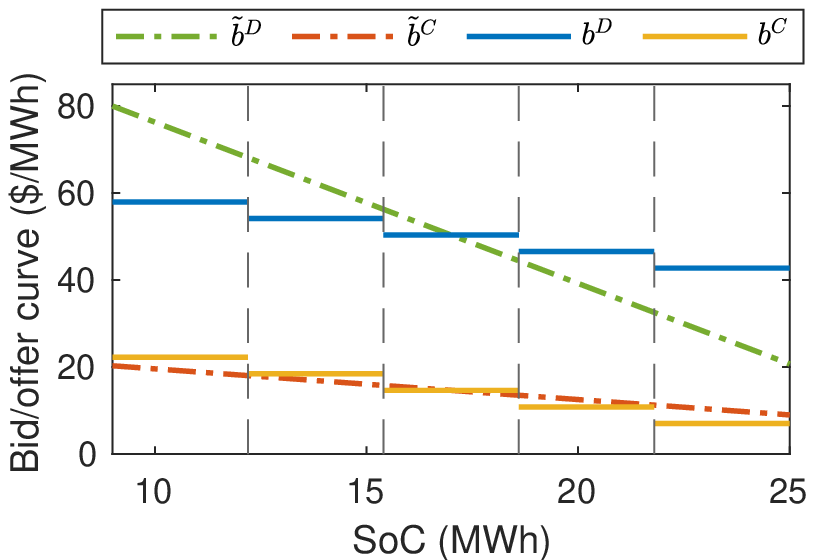}
\scalefig{0.24}\epsfbox{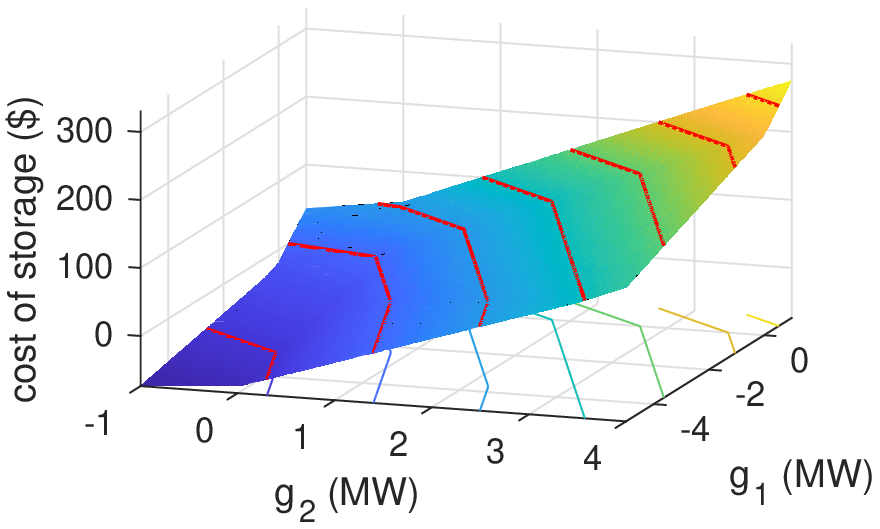}\scalefig{0.24}\epsfbox{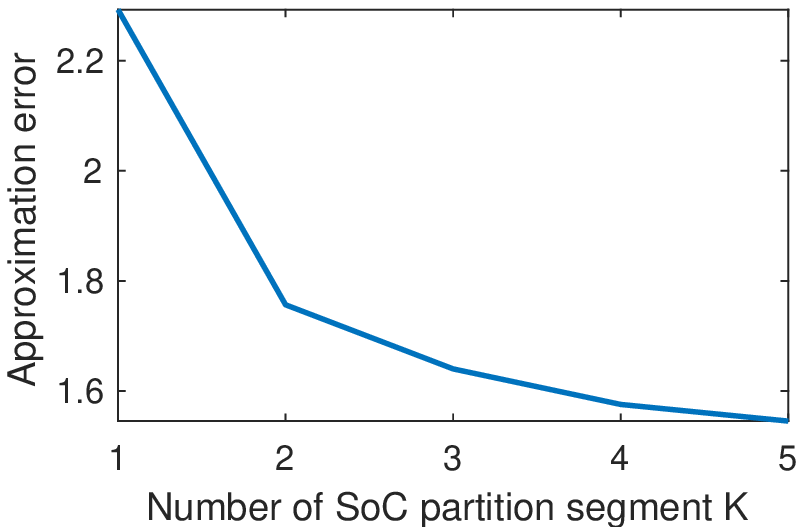}
% \scalefig{0.24}\epsfbox{figs/NonconvexSOCOver.eps}\scalefig{0.24}\epsfbox{figs/EqStepSizeOver.eps}
\end{psfrags}
\vspace{-0.5em}\caption{\scriptsize Top left: nonconvex true storage cost in 2-interval.  Top right: true SoC-dependent marginal cost and optimal EDCR approximation bids ($K=5$). Bottom left: optimal EDCR approximation cost  in 2-interval.  Bottom right: EDCR approximation error. }
\label{fig:SOCdependent}
\end{figure}
\vspace{-0.3cm}
Consider an ideal storage with the initial SoC at 15.5 MWh, $T=2$ and the original nonconvex multi-interval storage cost shown in the top left of Fig~\ref{fig:SOCdependent} with the axis label, $g_t=g^{\mbox{\tiny D}}_{t}-g^{\mbox{\tiny C}}_{t}, t=1,2,$ representing storage's net-producing power. The true SoC-dependent bids,  $\tilde{b}^{\mbox{\tiny C}}(e_t)$ and $\tilde{b}^{\mbox{\tiny D}}(e_t)$, are shown in Fig~\ref{fig:SOCdependent} (top right). From the EDCR approximation in (\ref{eq:OptEDCR_Bid}) with even SoC partitions\footnote{$\Ec_k$ satisfies $E_k=\underline{E}+\frac{(k-1)(\bar{E}-\underline{E})}{K}$ based on the SoC upper bound $\bar{E}=25$ MWh and lower bound $\underline{E}=9$MWh for all $k \in \{1,...,K\}$.} , we can approximate the true SoC-dependent bids and achieve the convex cost function shown in Fig.~\ref{fig:SOCdependent} (bottom left and top right). In this ideal storage which has $\eta^{\mbox{\tiny C}}=\eta^{\mbox{\tiny D}}=1$, the EDCR condition in Theorem~\ref{thm:Eq} decreased to $c^{\mbox{\tiny C}}_k-c^{\mbox{\tiny C}}_{k-1}=c^{\mbox{\tiny D}}_k-c^{\mbox{\tiny D}}_{k-1}, \forall k$ (shown in top right of Fig.~\ref{fig:SOCdependent}).

The bottom right part of Fig~\ref{fig:SOCdependent} illustrates the approximation error between the original SoC-dependent bids, $\tilde{b}^{\mbox{\tiny C}}(e_t)$ and $\tilde{b}^{\mbox{\tiny D}}(e_t)$, and the optimal EDCR approximation bids, $b^{\mbox{\tiny C}}(e_t)$ and $b^{\mbox{\tiny D}}(e_t)$. It is observed that, with more SoC partition segments, a smaller approximation error can be achieved.

\vspace{-0.1cm}
 \section{Conclusion}\label{sec:conclusion}
It's essential to remove non-convexities for a large-scale deployment of storage.  This paper convexifies the market clearing process by imposing a condition on the SoC-dependent bidding. We propose a sufficient condition---the equal decremental-cost ratio (EDCR) condition---to convexify the market clearing of multi-interval economic dispatch with SoC-dependent bids from merchant storage participants. And an optimal EDCR approximation method is proposed to compute the SoC-dependent bid from the true cost of  storage. 
\vspace{-0.1cm}

% \section*{Acknowledgement}
% The authors are grateful for the many discussions with   
{
\bibliographystyle{IEEEtran}
\bibliography{BIB}
}

\section*{Appendix}
\subsection{Proof of Theorem~\ref{thm:Eq}}

\underline{First, for equation (\ref{eq:ES_cost}), we prove} 
\beq
\begin{array}{lrl}
n&=&\underset{j\in\{1,...,K\}}{\rm arg~max}\{\alpha_j(s)-c^{\mbox{\tiny C}}_{j}\mathbf{1}^\intercal \mathbf{g}^{\mbox{\tiny C}}+c^{\mbox{\tiny D}}_{j}\mathbf{1}^\intercal \mathbf{g}^{\mbox{\tiny D}}\},\nn
\end{array}
\eeq
where $m$, $n$ and $r$ are respectively indexes for SoC-partitioned sets that has $e_{\mbox{\tiny 1}}=s\in \Ec_{m}$, $e_{\mbox{\tiny T+1}}\in \Ec_{n}$ and $e_{\mbox{\tiny T+2}}\in \Ec_{r}$.

By definition of notation $n$, we have $e_{\mbox{\tiny T+1}}\in[E_{n}, E_{n+1}]$. With $e_{\mbox{\tiny T+1}}=s+(\sum_{t=1}^{T}g^{\mbox{\tiny C}}_t)\eta^{\mbox{\tiny C}}-(\sum_{t=1}^{T}g^{\mbox{\tiny D}}_t)/\eta^{\mbox{\tiny D}}$, under Assumption~\ref{assume:single}, we have  $\forall p\in\{2,...,n\}, q\in\{n+1,...,K\} $ 
\beq
\begin{array}{clc}
& E_{n}\le s+\sum_{t=1}^{T}(g^{\mbox{\tiny C}}_t\eta^{\mbox{\tiny C}}-g^{\mbox{\tiny D}}_t/\eta^{\mbox{\tiny D}})\le E_{n+1},\\
&\Rightarrow  \begin{cases} E_{p}-s\le \sum_{t=1}^{T}(g^{\mbox{\tiny C}}_t\eta^{\mbox{\tiny C}}-g^{\mbox{\tiny D}}_t/\eta^{\mbox{\tiny D}}),& \\
\sum_{t=1}^{T}(g^{\mbox{\tiny C}}_t\eta^{\mbox{\tiny C}}-g^{\mbox{\tiny D}}_t/\eta^{\mbox{\tiny D}})\le E_{q}-s.& 
\end{cases}\\
&\Rightarrow \begin{cases} 
\Delta c^{\mbox{\tiny C}}_{p-1}(E_{p}-s)\le \Delta c^{\mbox{\tiny C}}_{p-1}\sum_{t=1}^{T}(g^{\mbox{\tiny C}}_t\eta^{\mbox{\tiny C}}-g^{\mbox{\tiny D}}_t/\eta^{\mbox{\tiny D}}),\nn \\
% ~~~~~~~~~~~~~~~~~~~~~~~~~~~~~~~~~~~~~~~~ \alpha\in\{2,...,s_{T+1}\}\\
\Delta c^{\mbox{\tiny C}}_{q-1}\sum_{t=1}^{T}(g^{\mbox{\tiny C}}_t\eta^{\mbox{\tiny C}}-g^{\mbox{\tiny D}}_t/\eta^{\mbox{\tiny D}})\le \Delta c^{\mbox{\tiny C}}_{q-1}(E_{q}-s ).
% ~~~~~~~~~~~~~~~~~~~~~~~~~~~~~~~~~~~~~~~~  \beta\in\{s_{T+1}+1,...,K\}
\end{cases}
\end{array}
\eeq

Known that  $\frac{c^{\mbox{\tiny C}}_k-c^{\mbox{\tiny C}}_{k-1}}{c^{\mbox{\tiny D}}_k-c^{\mbox{\tiny D}}_{k-1}}=\eta^{\mbox{\tiny C}}\eta^{\mbox{\tiny D}}, \forall k$, we have
\beq 
\begin{array}{clc}
&\Delta c^{\mbox{\tiny C}}_k\sum_{t=1}^{T}(g^{\mbox{\tiny C}}_t\eta^{\mbox{\tiny C}}-g^{\mbox{\tiny D}}_t/\eta^{\mbox{\tiny D}})=\sum_{t=1}^{T}(\Delta c^{\mbox{\tiny C}}_k g^{\mbox{\tiny C}}_t-\Delta c^{\mbox{\tiny D}}_kg^{\mbox{\tiny D}}_t)\eta^{\mbox{\tiny C}}\\
&~~=\sum_{t=1}^{T}\bigg((c^{\mbox{\tiny C}}_kg^{\mbox{\tiny C}}_t-c^{\mbox{\tiny D}}_kg^{\mbox{\tiny D}}_t)-(c^{\mbox{\tiny C}}_{k+1}g^{\mbox{\tiny C}}_t-c^{\mbox{\tiny D}}_{k+1}g^{\mbox{\tiny D}}_t)\bigg)\eta^{\mbox{\tiny C}}, \forall k.\nn
\end{array}
\eeq
So  $ \forall p\in\{2,...,n\}, \forall q \in \{n+1,...,K\}$,

$
\begin{cases}
\sum_{t=1}^{T}(c^{\mbox{\tiny D}}_{p-1}g^{\mbox{\tiny D}}_t-c^{\mbox{\tiny C}}_{p-1}g^{\mbox{\tiny C}}_t)+\frac{\Delta c^{\mbox{\tiny C}}_{p-1}}{\eta^{\mbox{\tiny C}}}(E_{p}-s)\\
~~~~~~~~~~~~~~~~~~~~~~~~~~~~~~~~\le\sum_{t=1}^{T}(c^{\mbox{\tiny D}}_{p}g^{\mbox{\tiny D}}_t-c^{\mbox{\tiny C}}_{p}g^{\mbox{\tiny C}}_t),\\
\sum_{t=1}^{T}(c^{\mbox{\tiny D}}_{q}g^{\mbox{\tiny D}}_t-c^{\mbox{\tiny C}}_{q}g^{\mbox{\tiny C}}_t)-\frac{\Delta c^{\mbox{\tiny C}}_{q-1}}{\eta^{\mbox{\tiny C}}}(E_{q}-s)\\
~~~~~~~~~~~~~~~~~~~~~~~~~~~~~~~~\le\sum_{t=1}^{T}(c^{\mbox{\tiny D}}_{q-1}g^{\mbox{\tiny D}}_t-c^{\mbox{\tiny C}}_{q-1}g^{\mbox{\tiny C}}_t).
\end{cases}$

$\Rightarrow \begin{cases}
-\sum_{k=1}^{p-2}\frac{c^{\mbox{\tiny C}}_k}{\eta^{\mbox{\tiny C}}}(E_{k+1}-E_{k})-\frac{c^{\mbox{\tiny C}}_{p-1}}{\eta^{\mbox{\tiny C}}}(s-E_{p-1})\\
~~~~+c^{\mbox{\tiny D}}_{p-1}\mathbf{1}^\intercal \mathbf{g}^{\mbox{\tiny D}}-c^{\mbox{\tiny C}}_{p-1}\mathbf{1}^\intercal \mathbf{g}^{\mbox{\tiny C}} \le c^{\mbox{\tiny D}}_{p}\mathbf{1}^\intercal \mathbf{g}^{\mbox{\tiny D}}-c^{\mbox{\tiny C}}_{p}\mathbf{1}^\intercal \mathbf{g}^{\mbox{\tiny C}}\\
~~~~-\sum_{k=1}^{p-1}\frac{c^{\mbox{\tiny C}}_k}{\eta^{\mbox{\tiny C}}}(E_{k+1}-E_{k})-\frac{c^{\mbox{\tiny C}}_{p}}{\eta^{\mbox{\tiny C}}}(s-E_{p}), \\
-\sum_{k=1}^{q-1}\frac{c^{\mbox{\tiny C}}_k}{\eta^{\mbox{\tiny C}}}(E_{k+1}-E_{k})-\frac{c^{\mbox{\tiny C}}_{q}}{\eta^{\mbox{\tiny C}}}(s-E_{q})\\
~~~~+c^{\mbox{\tiny D}}_{q}\mathbf{1}^\intercal \mathbf{g}^{\mbox{\tiny D}}-c^{\mbox{\tiny C}}_{q}\mathbf{1}^\intercal \mathbf{g}^{\mbox{\tiny C}} \le  c^{\mbox{\tiny D}}_{q-1 }\mathbf{1}^\intercal \mathbf{g}^{\mbox{\tiny D}}-c^{\mbox{\tiny C}}_{q-1 }\mathbf{1}^\intercal \mathbf{g}^{\mbox{\tiny C}}\\
~~~~-\sum_{k=1}^{q-2}\frac{c^{\mbox{\tiny C}}_k}{\eta^{\mbox{\tiny C}}}(E_{k+1}-E_{k})-\frac{c^{\mbox{\tiny C}}_{q-1 }}{\eta^{\mbox{\tiny C}}}(s-E_{q-1}).  \end{cases}$

$\Rightarrow \begin{cases}
\alpha_{p-1}(s)+c^{\mbox{\tiny D}}_{p-1}\mathbf{1}^\intercal \mathbf{g}^{\mbox{\tiny D}}-c^{\mbox{\tiny C}}_{p-1}\mathbf{1}^\intercal \mathbf{g}^{\mbox{\tiny C}} \\
~~~~~~~~~~~~~~~~\le c^{\mbox{\tiny D}}_{p}\mathbf{1}^\intercal \mathbf{g}^{\mbox{\tiny D}}-c^{\mbox{\tiny C}}_{p}\mathbf{1}^\intercal \mathbf{g}^{\mbox{\tiny C}}+\alpha_{p}(s), \\
\alpha_{q}(s)+c^{\mbox{\tiny D}}_{q}\mathbf{1}^\intercal \mathbf{g}^{\mbox{\tiny D}}-c^{\mbox{\tiny C}}_{q}\mathbf{1}^\intercal \mathbf{g}^{\mbox{\tiny C}}\\
~~~~~~~~~~~~~~~~ \le  c^{\mbox{\tiny D}}_{q-1 }\mathbf{1}^\intercal \mathbf{g}^{\mbox{\tiny D}}-c^{\mbox{\tiny C}}_{q-1 }\mathbf{1}^\intercal \mathbf{g}^{\mbox{\tiny C}}+\alpha_{q-1}(s).  \end{cases}$

In the last group of inequalities above, we use 

 $-\sum_{k=1}^{j-1}\frac{c^{\mbox{\tiny C}}_k}{\eta^{\mbox{\tiny C}}}(E_{k+1}-E_{k})-\frac{c^{\mbox{\tiny C}}_{j}}{\eta^{\mbox{\tiny C}}}(s-E_{j})+h(s)\\
 ~~~~ ~~~~=-\sum_{k=1}^{j-1}\frac{c^{\mbox{\tiny C}}_k}{\eta^{\mbox{\tiny C}}}(E_{k+1}-E_1+E_1-E_{k})\\
 ~~~~ ~~~~~~~~~~~~~~~~-\frac{c^{\mbox{\tiny C}}_{j}}{\eta^{\mbox{\tiny C}}}(s-E_1+E_1-E_{j})+h(s)\\
  ~~~~~~~~=-\sum_{k=1}^{j-1}\frac{\Delta c^{\mbox{\tiny C}}_k(E_{k+1}-E_1)}{\eta^{\mbox{\tiny C}}}-\frac{c^{\mbox{\tiny C}}_{j}(s-E_1)}{\eta^{\mbox{\tiny C}}}+h(s)\\
  ~~~~~~~~=\alpha_j(s).$

Therefore, we have
\beq
\begin{array}{lrl}
n=\underset{j}{\rm arg~max}\{\alpha_j(s)-c^{\mbox{\tiny C}}_{j}\mathbf{1}^\intercal \mathbf{g}^{\mbox{\tiny C}}+c^{\mbox{\tiny D}}_{j}\mathbf{1}^\intercal \mathbf{g}^{\mbox{\tiny D}}\},\nn
%-\sum_{k=1}^{j-1}\frac{c^{\mbox{\tiny C}}_k}{\eta^{\mbox{\tiny C}}}(E_{k+1}-E_{k})-\frac{c^{\mbox{\tiny C}}_j}{\eta^{\mbox{\tiny C}}}(s-E_{j})\\~~~~~~~~~~
\end{array}
\eeq
\beq\label{eq:ES_costComplete}
\begin{array}{lrl}
\sum_{t=1}^{T}f(g^{\mbox{\tiny C}}_{t}, g^{\mbox{\tiny D}}_{t};e_{t})=\underset{j\in\{1,...,K\}}{\rm max}\{\alpha_j(s)-c^{\mbox{\tiny C}}_{j}\mathbf{1}^\intercal \mathbf{g}^{\mbox{\tiny C}}+c^{\mbox{\tiny D}}_{j}\mathbf{1}^\intercal \mathbf{g}^{\mbox{\tiny D}}\}\\
~~~~~~~~=-c^{\mbox{\tiny C}}_{n}\mathbf{1}^\intercal \mathbf{g}^{\mbox{\tiny C}}+c^{\mbox{\tiny D}}_{n}\mathbf{1}^\intercal \mathbf{g}^{\mbox{\tiny D}}\\
~~~~~~~~~~~~~~+\begin{cases}
\sum_{k=m}^{n-1}\frac{-\Delta c^{\mbox{\tiny C}}_{k}}{\eta^{\mbox{\tiny C}}}(E_{k+1}-s), & n>m \\
0, & m=n  \\
\sum_{k=n+1}^{m}\eta^{\mbox{\tiny D}}\Delta c^{\mbox{\tiny D}}_{k-1}(E_{k}-s), & n<m
\end{cases}.
\end{array}
\eeq

 \underline{Next, we prove Theorem~\ref{thm:Eq} for all $T$ by induction.}

1) When $T=1$, the storage cost equals (\ref{eq:StageCost}), which is (\ref{eq:ES_costComplete}) with  $T=1$. From Assumption~\ref{assume:single}, (\ref{eq:ES_cost}) is convex. And this means Theorem~\ref{thm:Eq} is true at time $T=1$.

2) Assume Theorem~\ref{thm:Eq} is true at time $T$, and here we prove that Theorem~\ref{thm:Eq} is true at $T+1$. 

i. \underline{When $r=n$}, the total cost of storage at time $T+1$, i.e. $F(\gbf^{\mbox{\tiny C}}, \gbf^{\mbox{\tiny D}};s)=\sum_{t=1}^{T+1}f(g^{\mbox{\tiny C}}_{t}, g^{\mbox{\tiny D}}_{t};e_{t})$ equals to
\beq
\begin{array}{lrl}
\sum_{t=1}^{T}f(g^{\mbox{\tiny C}}_{t}, g^{\mbox{\tiny D}}_{t};e_{t})-c^{\mbox{\tiny C}}_{n}g^{\mbox{\tiny C}}_{\mbox{\tiny T+1}}+c^{\mbox{\tiny D}}_{n}g^{\mbox{\tiny D}}_{\mbox{\tiny T+1}}\\
~~~~=-c^{\mbox{\tiny C}}_{r}(\sum_{t=1}^{ T+1}g^{\mbox{\tiny C}}_t)+c^{\mbox{\tiny D}}_{r}(\sum_{t=1}^{ T+1}g^{\mbox{\tiny D}}_t)\\
~~~~~~~~+\begin{cases}
\sum_{k=m}^{r-1}\frac{-\Delta c^{\mbox{\tiny C}}_{k}}{\eta^{\mbox{\tiny C}}}(E_{k+1}-s), & r>m  \\
0, & r=m  \\
\sum_{k=r+1}^{m}\eta^{\mbox{\tiny D}}\Delta c^{\mbox{\tiny D}}_{k-1}(E_{k}-s), & r<m
\end{cases}.\nn
\end{array}
\eeq
So, Theorem~\ref{thm:Eq}  is true at time $T+1$, when $r=n$.

ii. \underline{When $r>n$} and $n>m$, we have $g^{\mbox{\tiny C}}_{\mbox{\tiny T+1}}>0, g^{\mbox{\tiny D}}_{\mbox{\tiny T+1}}=0$ from Lemma~\ref{lemma:bidSpread}. And from  (\ref{eq:StageCost}),  we have 
\beq
\begin{array}{l}
f^{\mbox{\tiny C}}(e_{\mbox{\tiny T+1}}, g_{\mbox{\tiny T+1}}^{\mbox{\tiny C}})=g^{\mbox{\tiny C}}_{\mbox{\tiny T+1}}c^{\mbox{\tiny C}}_{r}+\sum_{k=n}^{r-1}\frac{\Delta c^{\mbox{\tiny C}}_{k}}{\eta^{\mbox{\tiny C}}}(E_{k+1}-e_{\mbox{\tiny T+1}})\\
~~~~~~~~= (e_{\mbox{\tiny T+1}}+g^{\mbox{\tiny C}}_{\mbox{\tiny T+1}}\eta^{\mbox{\tiny C}}-E_{r})\frac{c^{\mbox{\tiny C}}_{r}}{\eta^{\mbox{\tiny C}}}+(E_{n+1}-e_{\mbox{\tiny T+1}})\frac{c^{\mbox{\tiny C}}_{n}}{\eta^{\mbox{\tiny C}}}\\
~~~~~~~~~~~~+\sum_{k=n+1}^{r-1}\frac{c^{\mbox{\tiny C}}_{k}}{\eta^{\mbox{\tiny C}}}(E_{k+1}-e_{\mbox{\tiny T+1}}+e_{\mbox{\tiny T+1}}-E_{k})\\
~~~~~~~~= (e_{\mbox{\tiny T+2}}-E_{r})\frac{c^{\mbox{\tiny C}}_{r}}{\eta^{\mbox{\tiny C}}}+(E_{n+1}-e_{\mbox{\tiny T+1}})\frac{c^{\mbox{\tiny C}}_{n}}{\eta^{\mbox{\tiny C}}}\\
~~~~~~~~~~~~+\sum_{k=n+1}^{r-1}\frac{c^{\mbox{\tiny C}}_{k}}{\eta^{\mbox{\tiny C}}}(E_{k+1}-E_{k}).\nn
\end{array}
\eeq
Additionally,  with $e_{t+1}=s+\sum_{t'=1}^{t}(g^{\mbox{\tiny C}}_{t'}\eta^{\mbox{\tiny C}}-g^{\mbox{\tiny D}}_{t'}/\eta^{\mbox{\tiny D}}), \forall t$, and $\frac{c^{\mbox{\tiny C}}_k-c^{\mbox{\tiny C}}_{k-1}}{c^{\mbox{\tiny D}}_k-c^{\mbox{\tiny D}}_{k-1}}=\eta^{\mbox{\tiny C}}\eta^{\mbox{\tiny D}}, \forall k$, the total cost of storage until $T+1$  is 
\beq
\begin{array}{lcl}
\sum_{t=1}^{T+1}f(g^{\mbox{\tiny C}}_{t}, g^{\mbox{\tiny D}}_{t};e_{t})=\sum_{t=1}^{T}f(g^{\mbox{\tiny C}}_{t}, g^{\mbox{\tiny D}}_{t};e_{t})-f^{\mbox{\tiny C}}(e_{\mbox{\tiny T+1}}, g_{\mbox{\tiny T+1}}^{\mbox{\tiny C}})\\
~~=\sum_{t=1}^{T}(c^{\mbox{\tiny D}}_{n}g^{\mbox{\tiny D}}_t-c^{\mbox{\tiny C}}_{n}g^{\mbox{\tiny C}}_t)+\sum_{k=m}^{n-1}\frac{-\Delta c^{\mbox{\tiny C}}_{k}}{\eta^{\mbox{\tiny C}}}(E_{k+1}-s)\\
~~~~~~-\frac{c^{\mbox{\tiny C}}_{r}}{\eta^{\mbox{\tiny C}}}(g^{\mbox{\tiny C}}_{\mbox{\tiny T+1}}\eta^{\mbox{\tiny C}}+s+\sum_{t=1}^{T}(g^{\mbox{\tiny C}}_t\eta^{\mbox{\tiny C}}-g^{\mbox{\tiny D}}_t/\eta^{\mbox{\tiny D}})-E_{r})\\
~~~~~~-\frac{c^{\mbox{\tiny C}}_{n}}{\eta^{\mbox{\tiny C}}}(-s+E_{n+1}-\sum_{t=1}^{T}(g^{\mbox{\tiny C}}_t\eta^{\mbox{\tiny C}}-g^{\mbox{\tiny D}}_t/\eta^{\mbox{\tiny D}}))\\
~~~~~~-\sum_{k=n+1}^{r-1}\frac{c^{\mbox{\tiny C}}_k}{\eta^{\mbox{\tiny C}}}(E_{k+1}-E_k)\nn\\
~~=(c^{\mbox{\tiny C}}_{n}-c^{\mbox{\tiny C}}_{n}-c^{\mbox{\tiny C}}_{r})(\sum_{t=1}^{T}g^{\mbox{\tiny C}}_t)+(c^{\mbox{\tiny D}}_{n}+\frac{c^{\mbox{\tiny C}}_{r}}{\eta^{\mbox{\tiny C}}\eta^{\mbox{\tiny D}}}-\frac{c^{\mbox{\tiny C}}_{n}}{\eta^{\mbox{\tiny C}}\eta^{\mbox{\tiny D}}})\\
~~~~~~(\sum_{t=1}^{T}g^{\mbox{\tiny D}}_t)+\sum_{k=m}^{n-1}\frac{-\Delta c^{\mbox{\tiny C}}_{k}}{\eta^{\mbox{\tiny C}}}(E_{k+1}-s)\\
~~~~~~-\sum_{k=n+1}^{r-1}\frac{c^{\mbox{\tiny C}}_k}{\eta^{\mbox{\tiny C}}}(E_{k+1}-s+s-E_k)-c^{\mbox{\tiny C}}_{r}g^{\mbox{\tiny C}}_{\mbox{\tiny T+1}} \\
~~~~~~-\frac{c^{\mbox{\tiny C}}_{n}}{\eta^{\mbox{\tiny C}}}(E_{n+1}-s)-\frac{c^{\mbox{\tiny C}}_{r}}{\eta^{\mbox{\tiny C}}}(s-E_{r})\\
~~=\sum_{t=1}^{T+1}(c^{\mbox{\tiny D}}_{r}g^{\mbox{\tiny D}}_t-c^{\mbox{\tiny C}}_{r}g^{\mbox{\tiny C}}_t)-\sum_{k=m}^{r-1}\frac{\Delta c^{\mbox{\tiny C}}_{k}}{\eta^{\mbox{\tiny C}}}(E_{k+1}-s).\nn
\end{array}
\eeq

Similarly, when $n=m$ and $n<m$, we can show that the total cost of storage at time $T+1$ is given by
\beq
\begin{array}{lrl}
\sum_{t=1}^{T+1}f(g^{\mbox{\tiny C}}_{t}, g^{\mbox{\tiny D}}_{t};e_{t})
%=\sum_{t=1}^{T+1}(c^{\mbox{\tiny D}}_{s_{T+2}}g^{\mbox{\tiny D}}_t-c^{\mbox{\tiny C}}_{s_{T+2}}g^{\mbox{\tiny C}}_t)\\
%~~~~+\begin{cases}
%\sum_{k=s_1}^{s_{T+2}-1}\frac{-\Delta c^{\mbox{\tiny C}}_{k}}{\eta^{\mbox{\tiny C}}}(E_{k+1}-e_1), & s_{T+2}>s_1  \\
%0, & s_{T+2}=s_1  \\
%\sum_{k=s_{T+2}+1}^{s_1}\eta^{\mbox{\tiny D}}\Delta c^{\mbox{\tiny D}}_{k-1}(E_{k}-e_1), & s_{T+2}<s_1
=\sum_{t=1}^{T+1}(c^{\mbox{\tiny D}}_{r}g^{\mbox{\tiny D}}_t-c^{\mbox{\tiny C}}_{r}g^{\mbox{\tiny C}}_t)\\
~~~~~~~~~~~~~~~~+\begin{cases}
\sum_{k=m}^{r-1}\frac{-\Delta c^{\mbox{\tiny C}}_{k}}{\eta^{\mbox{\tiny C}}}(E_{k+1}-s), & r>m  \\
0, & r=m  \\
\sum_{k=r+1}^{m}\eta^{\mbox{\tiny D}}\Delta c^{\mbox{\tiny D}}_{k-1}(E_{k}-s), & r<m
\end{cases}.\nn
\end{array}
\eeq

% \beq
% \begin{array}{lrl}
% \left\{\begin{matrix}
% -c^C_{s_{T+2}}(\sum_{t=1}^{T+1}g^C_t)+c^D_{s_{T+2}}(\sum_{t=1}^{T+1}g^D_t)+\sum_{k=s_1}^{s_{T+2}-1}(c^C_{k+1}-c^C_{k})(E_{k+1}-e_1), & if~s_{T+2}>s_1  \\
% -c^C_{s_{T+2}}(\sum_{t=1}^{T+1}g^C_t)+c^D_{s_{T+2}}(\sum_{t=1}^{T}g^D_t), & if~s_{T+2}=s_1  \\-c^C_{s_{T+2}}(\sum_{t=1}^{T+1}g^C_t)+c^D_{s_{T+2}}(\sum_{t=1}^{T+1}g^D_t)+\sum_{k=s_{T+2}+1}^{s_1}(c^C_{k-1}-c^C_{k})(E_{k}-e_1), & i+1f~s_{T+2}<s_1.
% \end{matrix}\right.
% \end{array}
% \eeq
    
iii.\underline{When $r<n$},  the same proof follows. Based on those operations that preserve convexity, the piecewise linear function (\ref{eq:ES_cost}) with Assumption~\ref{assume:single} is convex. So Theorem~\ref{thm:Eq}  is true at time $T+1$.

\subsection{Proof of Lemma~\ref{lemma:bidSpread}}
Proof: We prove Lemma~\ref{lemma:bidSpread} by contradiction. Assume that there exists an optimal solution with the simultaneous charging and discharging power, \ie $g^{\mbox{\tiny C}*}_{it}>0, g^{\mbox{\tiny D}*}_{it}>0$. After relaxing the constraint, $g^{\mbox{\tiny C}}_{it}g^{\mbox{\tiny D}}_{it}=0, \forall i, t,$, (\ref{eq:NONCVX}) is convex with a subdifferentiable objective when the  EDCR condition is satisfied. With the KKT conditions \cite[p. 281]{Rockafellar70convex}, there exist $ \kappa^{\mbox{\tiny C}}_i \in \frac{\partial}{\partial g^{\mbox{\tiny C}}_{it}}  f_i(g^{\mbox{\tiny C}*}_{it}, g^{\mbox{\tiny D}*}_{it}, e^*_{it}) $ and $ \kappa^{\mbox{\tiny D}}_i \in \frac{\partial}{\partial g^{\mbox{\tiny D}}_{it}}  f_i(g^{\mbox{\tiny C}*}_{it}, g^{\mbox{\tiny D}*}_{it}, e^*_{it}) $, satisfing   
\[
\begin{cases} 
 \kappa^{\mbox{\tiny C}}_i +\lambda^*_t-\Sbf(:,i)^\intercal\mubf^*[t]-\phi^*_{it}\eta^{\mbox{\tiny C}}_i-\underline{\rho}^{\mbox{\tiny C}*}_{it}+\bar{\rho}^{\mbox{\tiny C}*}_{it}=0& \\
 \kappa^{\mbox{\tiny D}}_i-\lambda^*_t+\Sbf(:,i)^\intercal\mubf^*[t]+\phi_{it}^*/\eta^{\mbox{\tiny D}}_i-\underline{\rho}^{\mbox{\tiny D}*}_{it}+\bar{\rho}^{\mbox{\tiny D}*}_{it}=0 & 
\end{cases}
\]
\beq
\label{eq:combKKT}
\Rightarrow  \frac{1}{\eta^{\mbox{\tiny C}}_i} \kappa^{\mbox{\tiny C}} _i+\kappa^{\mbox{\tiny D}}_i+\pi_{it}(\frac{1}{\eta^{\mbox{\tiny C}}_i}-\eta^{\mbox{\tiny D}}_i)+\frac{\bar{\rho}^{\mbox{\tiny C}*}_{it}}{\eta^{\mbox{\tiny C}}_i}+\bar{\rho}^{\mbox{\tiny C}*}_{it}\eta^{\mbox{\tiny D}}_i=0,
\eeq
%\[\Rightarrow 0 \in {\cal C}(g^{\mbox{\tiny C}*}_{it}, g^{\mbox{\tiny D}*}_{it};e^*_{it})+\lambda^*_t(\frac{1}{\eta^{\mbox{\tiny C}}}-\eta^{\mbox{\tiny D}})+\frac{\bar{\rho}^{\mbox{\tiny C}*}}{\eta^{\mbox{\tiny C}}}+\bar{\rho}^{\mbox{\tiny C}*}\eta^{\mbox{\tiny D}},\]
where $\pi_{it}$ is the LMP, $\bar{\rho}^{\mbox{\tiny C}*}_{it}\geq 0, \bar{\rho}^{\mbox{\tiny D}*}_{it}\geq 0, \frac{1}{\eta^{\mbox{\tiny C}}_i}\geq 1 \geq \eta^{\mbox{\tiny D}}_i$, and  we have $\underline{\rho}^{\mbox{\tiny C}*}_{it}=0, \underline{\rho}^{\mbox{\tiny D}*}_{it}=0$ from the complementary slackness conditions. 
%\[{\cal C}(g^{\mbox{\tiny C}*}_{it}, g^{\mbox{\tiny D}*}_{it};e^*_{it}):=\frac{-1}{\eta^{\mbox{\tiny C}}}\frac{\partial}{\partial g^{\mbox{\tiny C}}_{it}}  f_i(g^{\mbox{\tiny C}}_{it}, g^{\mbox{\tiny D}}_{it};e_{it}) +\eta^{\mbox{\tiny D}}\frac{\partial}{\partial g^{\mbox{\tiny D}}_{it}}  f_i(g^{\mbox{\tiny C}}_{it}, g^{\mbox{\tiny D}}_{it};e_{it}).\] 
%$\pi_{it}:=\lambda^*_t-\Sbf(:,i)^\intercal\mubf^*[t]$
The subgradient of  the storage cost function, $\frac{\partial}{\partial g^{\mbox{\tiny C}}_{it}}  f_i(g^{\mbox{\tiny C}}_{it}, g^{\mbox{\tiny D}}_{it}, e_{it}) $ and $\frac{\partial}{\partial g^{\mbox{\tiny D}}_{it}}  f_i(g^{\mbox{\tiny C}}_{it}, g^{\mbox{\tiny D}}_{it}, e_{it})$, can be respectively computed by%\footnote{Interior of set $\Ec_{ik}$ has $\mbox{Int}~\Ec_{ik}=(E_{ik}, E_{i(k+1)})$. } 
\[\frac{\partial}{\partial g^{\mbox{\tiny C}}_{it}}  f_i(g^{\mbox{\tiny C}}_{it}, g^{\mbox{\tiny D}}_{it}, e_{it}) =\begin{cases} \{-c^{\mbox{\tiny C}}_{ik}\}, & \mbox{if}~ g^{\mbox{\tiny C}}_{it}\in \mbox{Int}~\Ec_{ik}\\ %(E_{ik}, E_{i(k+1)})
[-c^{\mbox{\tiny C}}_{ik}, -c^{\mbox{\tiny C}}_{i(k+1)}], & \mbox{if}~ g^{\mbox{\tiny C}}_{it}=E_{i(k+1)} \end{cases},\footnote{The subgradient equals to $-c^{\mbox{\tiny C}}_{i1}$ if $g^{\mbox{\tiny C}}_{it}=E_{i1}$, and $-c^{\mbox{\tiny C}}_{i\mbox {\tiny K}}$ if $g^{\mbox{\tiny C}}_{it}=E_{i \mbox {\tiny (K+1)}}$.}\]

\[\frac{\partial}{\partial g^{\mbox{\tiny D}}_{it}}  f_i(g^{\mbox{\tiny C}}_{it}, g^{\mbox{\tiny D}}_{it}, e_{it})=\begin{cases} \{c^{\mbox{\tiny D}}_{ik}\}, & \mbox{if}~ g^{\mbox{\tiny D}}_{it}\in \mbox{Int}~\Ec_{ik}\\ %(E_{ik}, E_{i(k+1)})
[c^{\mbox{\tiny D}}_{i(k+1)}, c^{\mbox{\tiny D}}_{ik}], & \mbox{if}~ g^{\mbox{\tiny D}}_{it}=E_{i(k+1)} \end{cases}.\footnote{The subgradient equals to $c^{\mbox{\tiny D}}_{i1}$ if $g^{\mbox{\tiny D}}_{it}=E_{i1}$, and $c^{\mbox{\tiny D}}_{i\mbox {\tiny K}}$ if $g^{\mbox{\tiny D}}_{it}=E_{i \mbox {\tiny (K+1)}}$.}\]

So, under Assumption~\ref{assume:single}, we have $ \frac{1}{\eta^{\mbox{\tiny C}}_i} \kappa^{\mbox{\tiny C}}_i +\eta^{\mbox{\tiny D}}_i\kappa^{\mbox{\tiny D}}_i> 0$, which contradicts to the assumption that the LMP $\pi_{it}$ is nonnegative in equation (\ref{eq:combKKT}). %Therefore, with nonnegative LMP, we won't have  the simultaneously charging and discharging behavior for the optimal solution.

\edoc

\end{document}